\documentclass[12pt,preprint]{elsarticle}
\usepackage[active]{srcltx} 
\usepackage{hyperref}
\usepackage{graphics}
\usepackage{epsfig}
\usepackage{amsmath}
\usepackage{mathtools}
\usepackage[english]{babel}
\usepackage{graphicx}
\usepackage{url}
 \usepackage{footnote}

\RequirePackage{lineno}
\makeatletter
\def\ps@pprintTitle{%
 \let\@oddhead\@empty
 \let\@evenhead\@empty
 \def\@oddfoot{}%
 \let\@evenfoot\@oddfoot}
\makeatother

\begin{document}

\title{QCD under extreme conditions}  

\begin{frontmatter}

\author[EMMI,HEIDELBERG]{Peter Braun-Munzinger}
\author[GSI,NNRC]{Anar Rustamov}
\author[HEIDELBERG]{Johanna Stachel}

\address[EMMI]{Extreme Matter Institute EMMI, GSI, 64291 Darmstadt, Germany}
\address[HEIDELBERG]{Physikalisches Institut, Universit\"{a}t Heidelberg, 69120 Heidelberg, Germany}
\address[GSI]{GSI Helmholtzzentrum f\"ur Schwerionenforschung GmbH, 64291 Darmstadt, Germany
}
\address[NNRC]{National Nuclear Research Center, AZ1000 Baku, Azerbaijan}

\begin{abstract}
In nucleus-nucleus collisions at relativistic energies a new kind of matter is 
created, the Quark-Gluon Plasma (QGP). The phase diagram of such  matter
and the chemical freeze-out points will be presented in connection to the pseudo-critical temperature for the chiral cross over transition. The role of conserved charge fluctuations to give experimental access to the nature of the chiral phase transition will be summarized in terms of the relation to lattice QCD and the current experimental data. The QGP can be characterized as a nearly ideal liquid expanding hydrodynamically and the experimental data allow to extract transport parameters such as the bulk and shear viscosities. The energy loss of partons in the QGP  probes the high parton density of the medium. The role of quarkonia and open charm hadrons  as a probe of deconfinement and hadronization form the final topic.
\vspace{10pt}
\end{abstract}

\end{frontmatter}

\section*{}
\begin{center}
\emph{Article to appear in a special EPJC Volume in celebration of  \\ '50 Years of Quantum Chromodynamics'.}
\end{center}
\par\noindent\rule{\textwidth}{0.4pt}
\vspace{4pt}

The infrared slavery and asymptotic freedom properties of QCD, discussed in previous sections, form the theoretical basis that strongly interacting matter at finite temperature and/or density exists in different thermodynamic phases. This was realized~\cite{Collins:1974ky,Cabibbo:1975ig} already short\-ly after these properties of QCD were introduced. The term quark-gluon plasma was coined soon after by Shuryak~\cite{Shuryak:1977ut} for the high temperature/density phase where confinement is lifted and a global symmetry of QCD, the chiral symmetry, is restored. The first lattice QCD (lQCD) calculations of the equation of state were performed soon thereafter~\cite{Engels:1980ty}. Already in early lQCD calculations a close link between deconfinement and restoration of chiral symmetry was found~\cite{Kogut:1982rt}. 

\begin{figure}[!htb]
    \centering
    \includegraphics[width=.76\linewidth,clip=true]{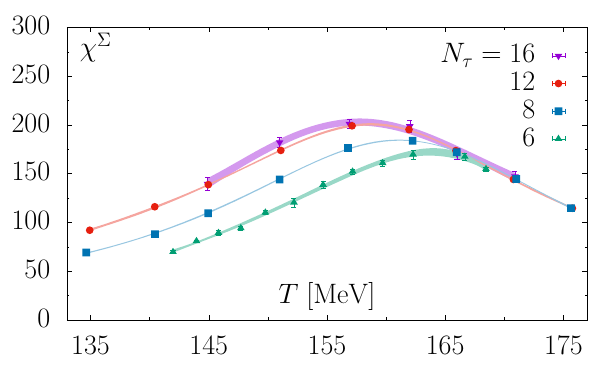}
    \caption{Susceptibility of the chiral u,d- and s-quark condensate as a function of temperature computed in 2+1 flavor lQCD (Fig. from~\cite{HotQCD:2018pds}).}
    \label{fig:chiral}
\end{figure}

\begin{figure}[!htb]
    \centering
    \includegraphics[width=.75\linewidth,clip=true]{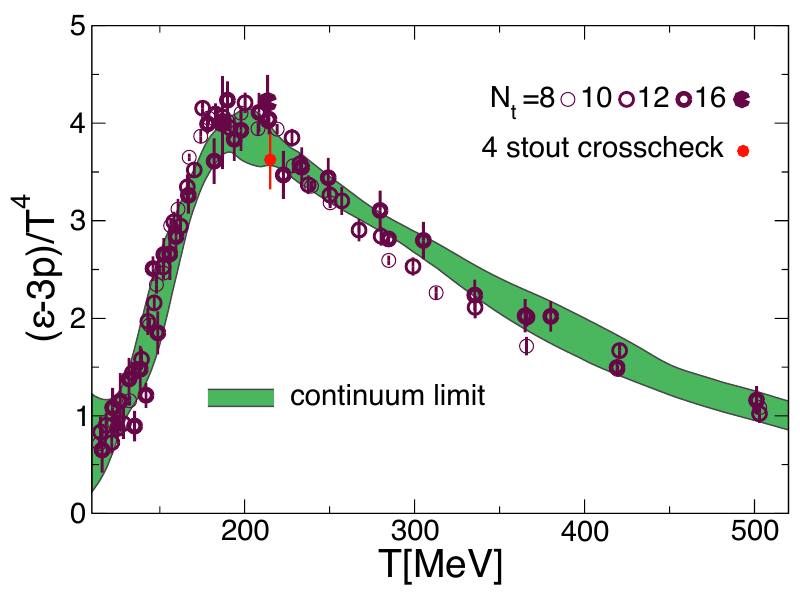}
    \caption{The interaction measure or trace anomaly normalized to the fourth power of the temperature as a function of temperature, computed in 2+1 flavor lQCD (Fig. from~\cite{Borsanyi:2013bia}).}
    \label{fig:trace}
\end{figure}

For deconfinement there is an order parameter for the phase transition, the so-called Polyakov loop, in the limit without dynamical quarks. For chiral symmetry restoration the chiral condensate $\langle  \bar\psi \psi \rangle$ forms an order parameter for vanishing but also for finite quark masses. Indeed, recent numerical lQCD calculations~\cite{HotQCD:2019xnw} provide, in the limit of massless u and d quarks,  strong indications for a genuine second-order chiral transition between a hadron gas and a QGP at a critical temperature of $T_{c}\approx132^{+3}_{-6}$ MeV. For realistic u,d,s-quark masses, chiral symmetry is restored in a crossover transition at vanishing net-baryon density and a precisely determined  pseudo-critical temperature of $T_{pc}$ = 156.5 $\pm 1.5$ MeV ~\cite{HotQCD:2018pds}. Consistent with this result, a transition temperature of 158.0 $\pm$ 0.6 MeV was recently reported in~\cite{Borsanyi:2020fev}. This pseudo-critical temperature is found as a maximum in the susceptibility (derivative with respect to mass) of the chiral condensate as displayed in Fig~\ref{fig:chiral}. Contrary to early ideas, the system remains strongly coupled over a rather large temperature range above $T_{pc}$. This is reflected in the interaction measure computed in lQCD as the difference between the energy density and three times the pressure, $I = \epsilon - 3 P$, which by definition vanishes for an ideal gas of massless quarks and gluons. Fig.~\ref{fig:trace} shows that this interaction measure, normalized to the fourth power of the temperature, peaks at about 20\% above $T_{pc}$ and falls off only slowly towards higher temperature values.

The lQCD calculations have been extended into the region of finite net baryon density quantified by a baryon chemical potential $\mu_B$~\cite{HotQCD:2018pds,Borsanyi:2020fev}. Current lQCD expansion techniques are valid in the regime of $\mu_B/T\leq 3$. The so obtained line of pseudo-critical temperatures is shown in the QCD phase diagram displayed in Fig~\ref{fig:phase_diagram} below. Because of the sign problem, the lQCD technique cannot be applied for still larger values of $\mu_B$, see e.g. ~\cite{Gattringer:2016kco}, and one has to resort to models of QCD for theoretical guidance in the high net baryon density region. 

Experimentally, this regime of the QCD phase transitions is accessible by investigating collisions of heavy nuclei at high energy. It was conjectured already in~\cite{Shuryak:1978ij} that, in such hadronic collisions, after some time local thermal equilibrium is established and all properties of the system (fireball) are determined by a single parameter, the temperature $T$, depending on time and spatial coordinates. This is exactly the regime probed by collisions of nuclei at the Large Hadron Collider (LHC), as will be outlined in the following. The region of finite to large $\mu_B$ is accessed by nuclear collisions at lower energies.

In the following, we describe the experimental efforts, principally at the LHC and at RHIC (Relativistic Heavy Ion Collider), to provide from analysis of relativistic nuclear collision data quantitative information on the QCD phase diagram by studying hadron production as a function of the nucleon-nucleon center of mass energy $\sqrt{s_{\mathrm{NN}}}$. We can only touch a small fraction of the physics of the quark-gluon plasma (QGP) in this brief review. Excellent summaries of the many other interesting topics can be found in recent review articles ~\cite{Jacak:2012dx,Muller:2012zq,Braun-Munzinger:2015hba,Busza:2018rrf}. 

\begin{figure}[!htb]
    \centering
    \includegraphics[width=.7\linewidth,clip=true]{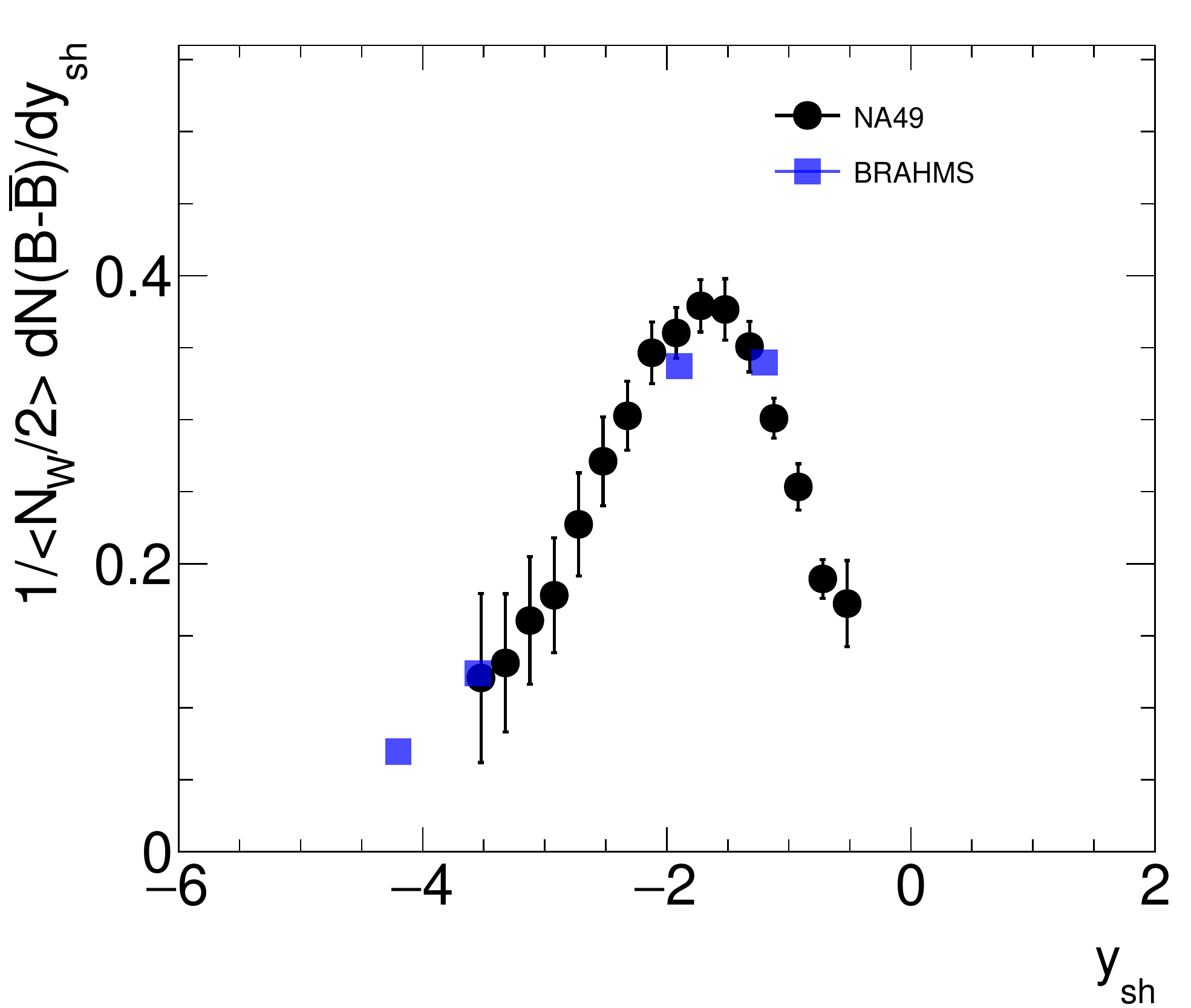}
    \caption{Normalized net-baryon rapidity densities for $\sqrt{s_{NN}} =$~ 17.3 and 62.4 GeV after subtracting the corresponding target contributions using the limiting fragmentation concept. Here $y_{sh}= y - y_{b}$ with $y_{b}$ the beam rapidity.}
    \label{fig:stopping}
\end{figure}

In the early phase of the collision, the incoming nuclei lose a large fraction of their energy leading to the creation of a hot fireball characterized by an energy density $\epsilon$ and a temperature $T$.  This stopping is characterized by the average rapidity shift of the incident nucleons, with $\Delta y$ = -ln($E/E_0$). Quantitative information is contained in the experimentally measured net-proton rapidity distributions (i.e. the difference between proton and anti-proton rapidity distributions). These distributions are presented for different collision energies from the SPS to RHIC energy range in~\cite{Braun-Munzinger:2020jbk}. There it can be seen that the rapidity shift saturates at approximately two units from $\sqrt{s_{NN}}\approx$ 17.3 GeV upwards, implying a fractional energy loss of $1-{\rm exp}(-\Delta y) \approx 86\%$. In fact, the same rapidity shift was already determined for p--nucleus collisions at Fermilab for 200 GeV/c proton momentum~\cite{Abe:1988hq}. With increasing collision energy, the target and projectile rapidity ranges are well separated, leaving at central rapidity a net-baryon depleted or even free high energy density region. Fig.~\ref{fig:stopping} shows the distribution of slowed down beam nucleons, after subtracting the tail of the target distribution and plotted against rapidity minus beam rapidity. It is apparent that up to $\sqrt{s_{NN}}$ = 62.4 GeV the concept of limiting fragmentation~\cite{Benecke:1969sh} is well realized. At higher energies, this rapidity region is very hard to reach experimentally for identified particles.

The rapidity shift of the incident nucleons leads to high energy densities at central rapidity, i.e., in the center of the fireball. These initial energy densities can be estimated, after fixing the kinetic equilibration time scale $\tau_0$, using the Bjorken model~\cite{Bjorken:1982qr}:
\begin{equation}
\label{eq-Bj}
    \epsilon_{BJ} = \frac{1}{A\tau_{0}}\frac{d\eta}{dy}\frac{dE_{T}}{d\eta},
\end{equation}
where $A=\pi r^{2}$ is the overlap area of two nuclei. Eq.~\ref{eq-Bj} is evaluated at a time $\tau_{0}$ = 1 fm and the resulting energy densities are displayed in Table~\ref{tab:Edensity} for central Au--Au and Pb--Pb collisions. For central Pb--Pb collisions ($A =$ 150 fm$^2$) at $\sqrt{s_{NN}}$ = 2.76 TeV this yields an energy density of about 14 GeV/${\rm fm^3}$~\cite{CMS:2012krf}, more than a factor of 30 above the critical energy density for the chiral phase transition as determined in lQCD calculations. In fact, for all collision energies shown the initial energy density significally exceeds the energy density computed in lQCD at the pseudo-critical temperature, indicating that  the matter in the fireball is to be described with quark and gluon degrees of freedom rather than as hadronic matter. The corresponding initial tem\-pe\-ra\-tures can be computed using the energy density of a gas of quarks and gluons with two quark flavors, $\epsilon = 37\frac{\pi^{2}}{30}T^{4}$, yielding $T\approx$ 307 MeV. Temperature values for lower collision energies are also quoted in the Table\footnote{The values reported in the table are all for vanishing chemical potentials. We have evaluated the differences if one assumes values for chemical potentials as determined at chemical freeze-out, see below. The resulting temperature values differ by less than 5\%  from those reported in Table~\ref{tab:Edensity}. Owing to the proportionality of energy density to the fourth power of temperature, inclusion of a bag pressure only mildly changes the calculated temperature values.}.
It can be seen that already at AGS energies the estimated values of $\epsilon$ and $T$ are significantly above the values for the chiral cross over transition.

\begin{table}[htb]
\centering
\begin{tabular}{| c | c | c | c | c |}
 \hline
 & $\sqrt{s_{NN}}$  & $dE_{t}/d\eta$ & $\epsilon_{BJ}$ & T \\ [2ex]
 & [GeV] & [GeV] & [GeV/$fm^{3}$] & [GeV] \\ 
 \hline
AGS & 4.8 & 200 & 1.9 & 0.180 \\  [1ex]
 \hline
SPS & 17.2 & 400 & 3.5 & 0.212 \\ [1ex]
\hline
RHIC & 200 & 600 & 5.5 & 0.239 \\ [1ex]
\hline
LHC & 2760 & 2000 & 14.5 & 0.307 \\ [1ex]
\hline
\end{tabular}
\caption{Collision energy, measured transverse energy pseudo-rapidity density at mid-rapidity~\cite{E814E877:1993rlr,WA98:2000mvt,PHENIX:2015tbb,CMS:2012krf}, energy density, and initial temperature estimated as described in the text for central Pb--Pb and Au--Au collisions at different accelerators.}
\label{tab:Edensity}
\end{table}

Depending on energy, collisions of heavy ions populate different regimes falling into two categories: (i) the stopping or high baryon density region reached at $\sqrt{s_{NN}} \approx$ 3-20 GeV and (ii) the transparency or baryon-free region reached at $\sqrt{s_{NN}} > $100 GeV. The net-baryon-free QGP presumably existed in the early Universe after the electro-weak phase transition and up to a few microseconds after the Big Bang\footnote{In the QGP of the early universe, particles interacting via the strong and  electro-weak force are part of the system, while an accelerator-made QGP only contains strongly interacting particles. }. On the other hand, a baryon-rich QGP could be populated in neutron star mergers or could exist, at very low temperatures, in the center of neutron stars\cite{Bauswein:2018bma,Baym:2019iky}.

For the system considered to come into local thermal equilibrium and, more importantly, for the development of a phase transition, the presence of interactions is necessary. In fact, close to the phase transition, the system has to be strongly coupled. As mentioned above, quarks and gluons under the extreme conditions reached in nuclear collisions are indeed strongly coupled. The large values of the interaction measure from lQCD calculations $(\epsilon - 3P)/T^4$, introduced above in Fig.~\ref{fig:trace}, lend support to the strong coupling scenario. Further, the energy and entropy densities $\epsilon/T^{4}$ and  $s/T^{3}$, as calculated in lQCD,  fall significantly short (by about 20 \%) of the Stefan-Boltzmann limit for an ideal gas of quarks and gluons up to a few times the pseudo-critical temperature. The conclusion about a strongly coupled QGP close to $T_{pc}$ also follows from experimental results at the colliders, and even at the SPS, on the coefficients of azimuthal anisotropies of hadron distributions in combination with a viscous hydrodynamic description.

\begin{figure}[!htb]
    \centering
    \includegraphics[width=.8\linewidth,clip=true]{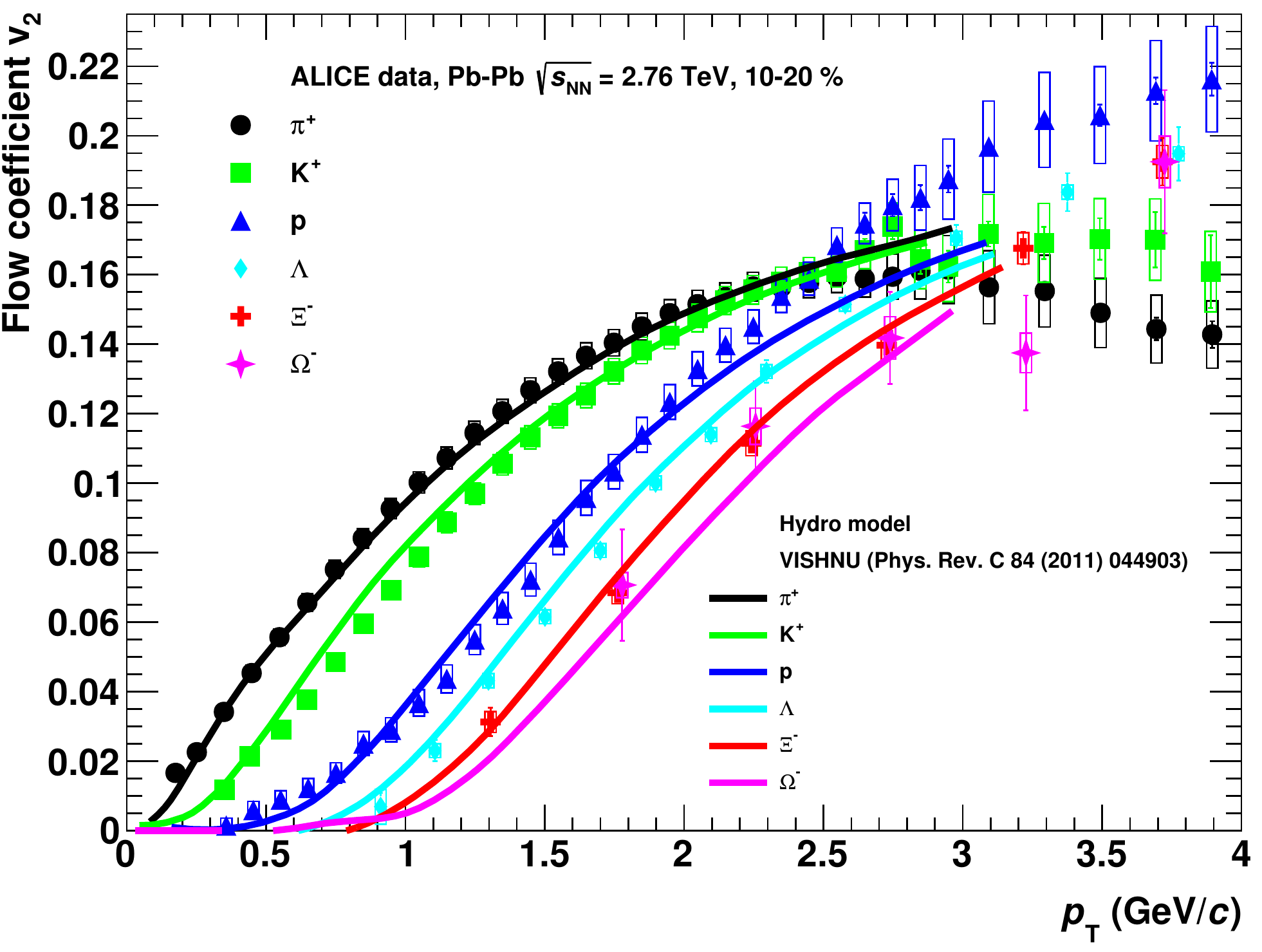}
    \caption{Elliptic flow coefficient $v_2$ for identified hadrons as a function of transverse momentum measured by ALICE and compared to results from viscous hydrodynamics calculations~\cite{Shen:2011eg}.}
    \label{fig:v2}
\end{figure}

\begin{figure}[!htb]
    \centering
    \includegraphics[width=1.0\linewidth,clip=true]{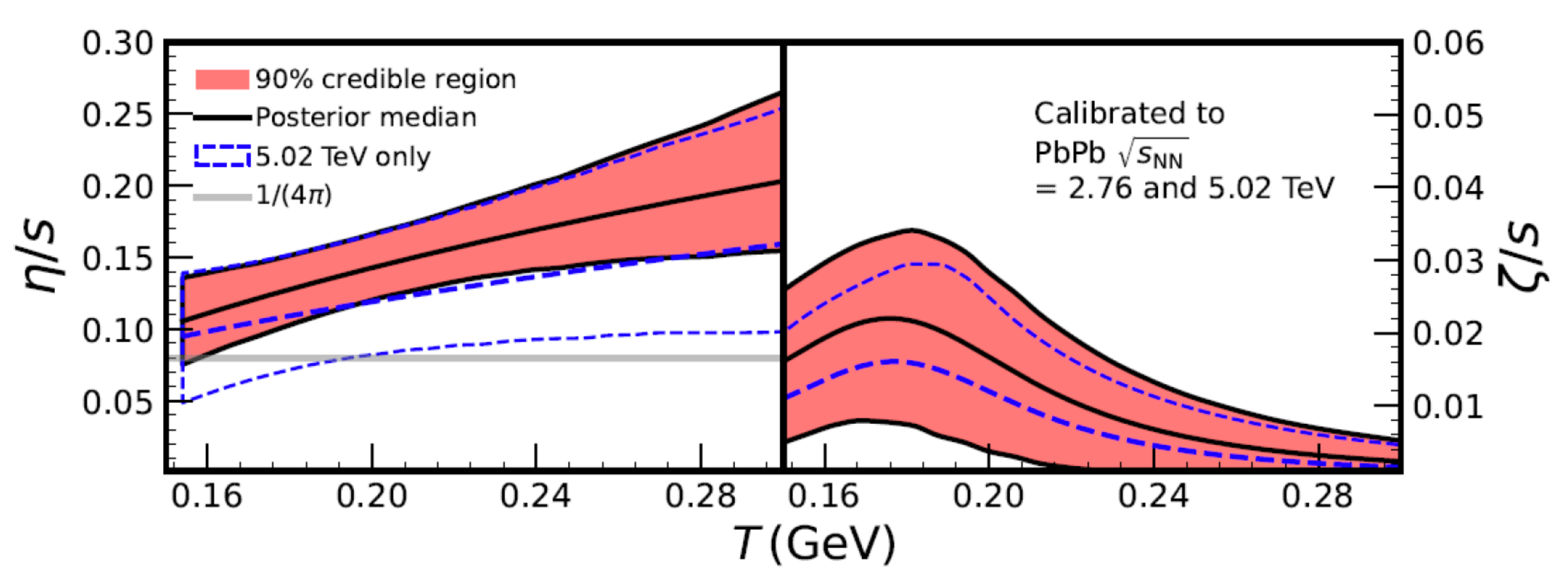}
    \caption{Temperature dependence of the shear (left panel) and bulk (right panel) viscosity to entropy density ratios. Figure taken from~\cite{Parkkila:2021yha}. }
    \label{fig:EtaOverS}
\end{figure}

For non-central nuclear collisions the distributions in transverse momentum $p_T$ of hadrons exhibit modulations with respect to the azimuthal angle $\phi$ in the reaction plane. These anisotropies  can be characterized by $p_T$ dependent Fourier coefficients. The dominant term is the 2nd order Fourier coefficient $v_2$, also called the elliptic flow coefficient. This modulation has been predicted to arise from the anisotropy of the gradient of the pressure P in the early phase of the collision due to the geometry of the nuclear overlap region, leading to correspondingly larger expansion velocities in the reaction plane and hence large $v_2$ coefficients.

 The strength of the coupling can be quantified by introducing transport parameters for the QGP such as the shear viscosity $\eta$, which is related to the mean free path of quarks and gluons inside the QGP, and the bulk viscosity $\zeta$, with its connection to QGP expansion dynamics and speed of sound. The smaller the transport coefficients the stronger the coupling. Larger values of the shear viscosity, e.g.,  suppress the magnitude of the elliptic flow. 
 
 For a strongly coupled system with small enough values of mean free path (comparable to or lower than the corresponding de Broglie wavelength of particles), treatment as a fluid is more appropriate. One then describes its properties by solving hydrodynamic equations. The shear viscosity enters the hydrodynamic equations as $\eta/(\epsilon + P) = \eta/(Ts)$, hence the quantity characterizing the medium is $\eta/s$. By comparing flow observables measured in experiments at RHIC~\cite{PHENIX:2003qra, STAR:2004jwm} and LHC~\cite{ALICE:2010suc} to the corresponding calculations in viscous hydrodynamics, accompanied with converting the fluid into thermal distributions of hadrons at the freeze-out hyper-surface, remarkably low values for $\eta/s$ are obtained. Fig.~\ref{fig:v2} shows as an example the elliptic flow coefficients $v_2$ for different identified hadrons at the LHC. A mass ordering characteristic for a hydrodynamically expanding medium is observed very clearly. And indeed, the mass ordering and its $p_T$ dependence are described quantitatively by a relativistic viscous hydrodynamic calculation~\cite{Shen:2011eg} as indicated by the lines in Fig.~\ref{fig:v2} employing a small ratio of $\eta/s$. 
  
 In fact, a lower bound of $\eta/s = 1/(4\pi)$ (in units of $\hbar = k_{B} = 1$) can be obtained for a large class of strongly coupled field theories from the quantum mechanical uncertainty principle~\cite{Danielewicz:1984ww} and using the AdS/CFT correspondence~\cite{Kovtun:2004de,Baier:2007ix, York:2008rr}. Recently, the values and the temperature dependence of the 
 shear and bulk viscosities employed in hydrodynamic codes were extracted by fitting spectra and azimuthal anisotropies of hadrons measured at the LHC and RHIC using Bayesian estimation methods~\cite{Bernhard:2019bmu, Parkkila:2021yha}. An example is shown in Fig.~\ref{fig:EtaOverS}. Inspection of this figure indicates that, at $T_{pc}$, the estimated value of $\eta/s$ is close to the lower bound of $1/(4\pi$) , indicating that the observed matter is a nearly perfect fluid. Above the transition temperature, the extracted band for $\eta/s$ is rising, reflecting a weakening of the coupling, although even at twice $T_{pc}$ the medium is still strongly coupled.
 On the other hand, as presented in Fig.~\ref{fig:trace}, near the phase transition the lQCD results exhibit a maximum in the interaction measure, which is an indication for interactions in the system. In the hydrodynamic calculations the breaking of scale invariance is accounted for by introducing a bulk viscosity $\zeta$ along with the shear viscosity. While increasing sheer viscosity reduces the momentum anisotropy, hence lowering the elliptic flow coefficients, the bulk viscosity reduces the overall rate of the radial expansion. 
 The right panel of Fig.~\ref{fig:EtaOverS} shows the temperature dependence of $\zeta/s$, which exhibits a peak just above the transition temperature~\cite{Parkkila:2021yha}. This location of the maximum is consistent with the temperature dependence of the interaction measure from lQCD. 
 
Important information on the structure of the QGP is also obtained by studying the interaction of high-momentum partons with the thermalized quarks and gluons in the QGP. A strongly coupled QGP is opaque to high momentum partons, leading to the phenomenon of 'jet quenching'~\cite{Busza:2018rrf}. In fact, the theoretical foundation for strong jet quenching by QCD bremsstrahlung was laid by~\cite{Baier:1996kr}. There it was shown that, for sufficiently energetic quarks and gluons, such that the radiation does not decohere, the radiative energy loss scales quadratically with the length traversed, leading to very large values. An important experimental observable linked to jet quenching is the observed suppression ('quenching') of high-momentum hadrons in central nuclear collisions at high collision energy. This suppression is quantified by the p$_T$ dependence of the ratio R$_{AA}$ of inclusive hadron production in collisions of nuclei with mass number A to that in proton-proton collisions, taking into account the collision geometry by scaling to the number of binary collisions~\cite{Busza:2018rrf}.

\begin{figure}[!htb]
   \centering
    \hspace{-1.3cm}
    \includegraphics[width=.72\linewidth,clip=true]{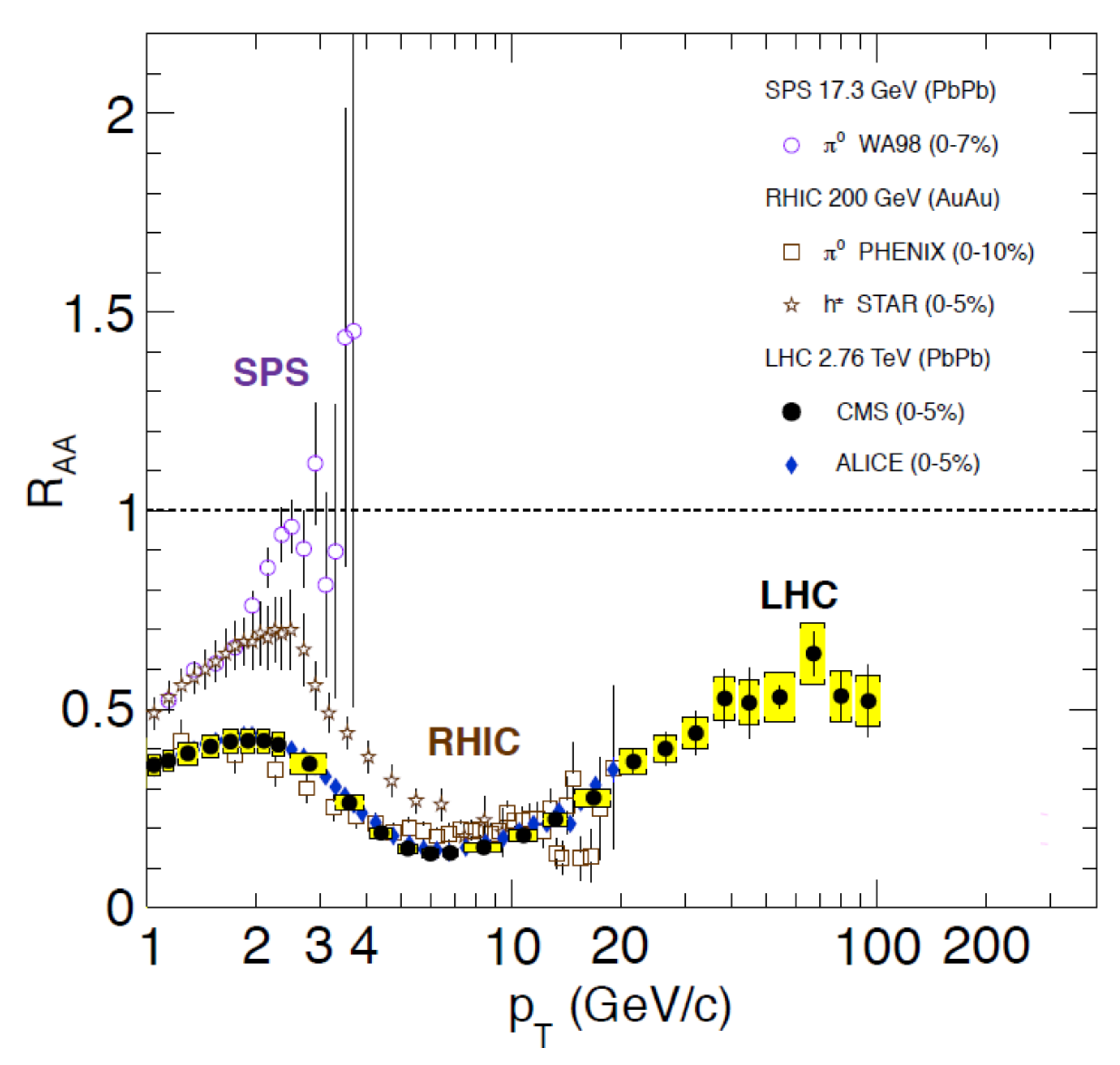}
    \caption{Evolution of the transverse momentum dependence of R$_{AA}$ for leading particles for central nuclear collisions with collision energy\cite{CMS:2012aa}.}
    \label{fig:jet_RAA}
\end{figure}

In Fig.~\ref{fig:jet_RAA} we present the evolution with cm energy of the transverse momentum dependence of R$_{AA}$ for leading particles as obtained from measurements at the SPS, RHIC, and LHC accelerators. Note that, by construction,  R$_{AA}$ = 1 for hard binary collisions in the absence of nuclear effect such as jet quenching. At very low p$_T$ one observes R$_{AA}$ values less than unity and increasing with p$_T$ since soft particle production scales with the number of participating nucleons and not the number of binary collisions. For RHIC and LHC energies the jet quenching is born out by a decreasing trend observed for p$_T > 2.5$ GeV/c reaching a broad minimum near p$_T = 7$ GeV/c of R$_{AA} = 0.1 - 0.2$: high momentum hadrons are quenched by about a factor of 5 or more. At LHC energies R$_{AA}$ increases again for higher p$_T$ values until a plateau is reached above p$_T \approx 100$ GeV/c. Measurements for fully reconstructed jets have been performed by the ATLAS  collaboration. The results demonstrate~\cite{ATLAS:2018gwx} that the quenching by about a factor of 2 persists to the highest available jet p$_T$ values of 1 TeV/c.

The data on jet quenching have been modeled in terms of elastic and inelastic collisions of partons in the dense QGP, taking into account important coherence effects~\cite{Zapp:2008gi,Armesto:2011ht}. For a recent summary see ~\cite{JETSCAPE:2021ehl} and ref. cited there.

\begin{figure}[!htb]
    \centering
    \includegraphics[angle=0,width=.75\linewidth,clip=true]{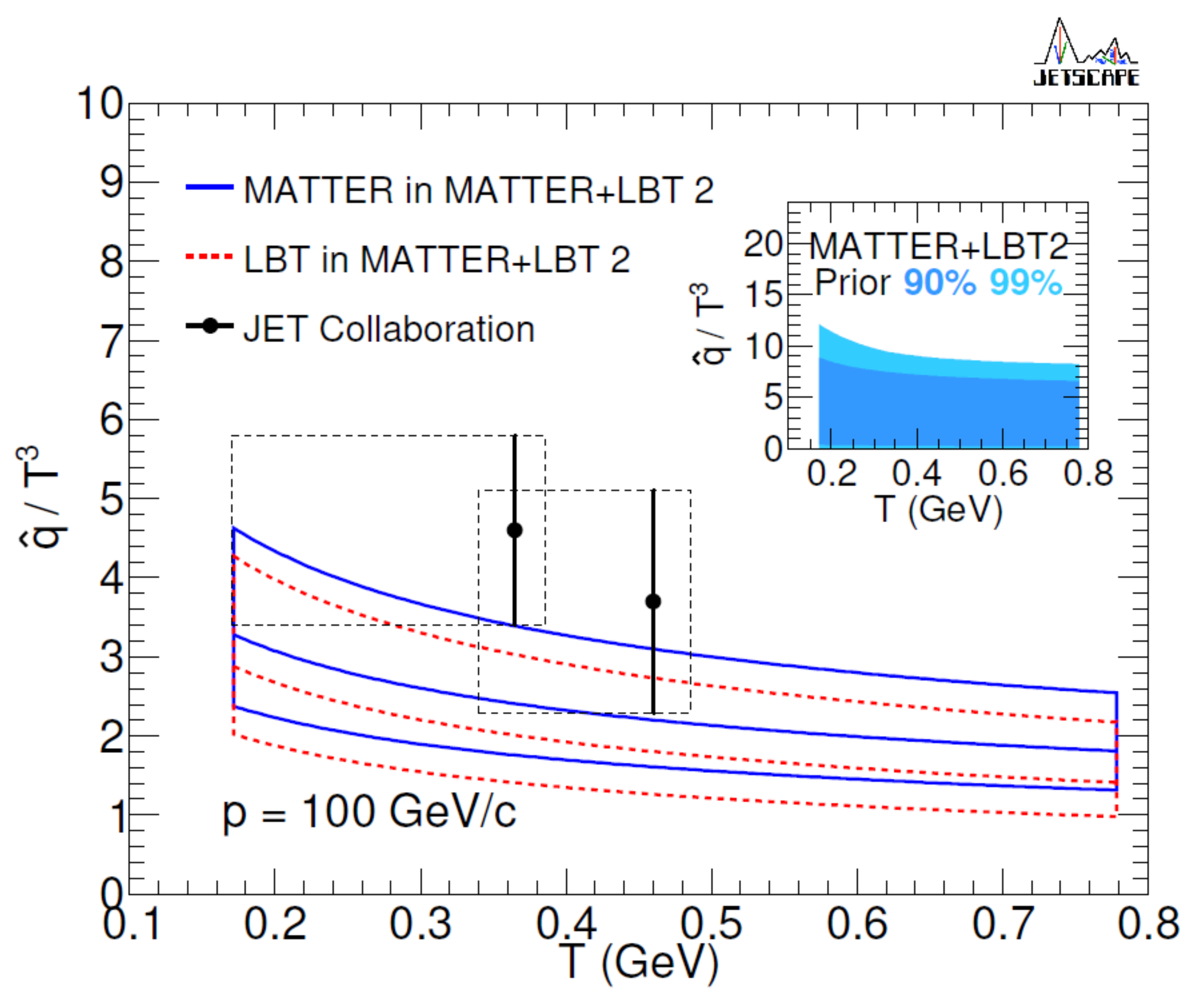}
    \caption{QGP jet transport parameter $\hat{q}/T^3$ obtained by the JETSCAPE collaboration\cite{JETSCAPE:2021ehl}.}
    \label{fig:qhat}
\end{figure}

To model experimental data with QCD-based jet quenching calculations one has to take into account that the jet is created as a product of an initial hard parton-parton collision with large momentum transfer Q. That implies that the parton initiating the jet is highly virtual. The magnitude of its 4-momentum Q as reflected in the total jet energy E can be hundreds of GeV (or even a few TeV at the LHC) while, for a real parton, Q$^2\approx 0$. The highly virtual parton will evolve into a parton shower which eventually hadronizes to form a collimated jet of hadrons. During the entire evolution the highly virtual initial parton and the parton shower components lose energy by interactions with the QGP constituents, resulting in a medium-modification of the entire parton fragmentation pattern, i.e. the jet~\cite{Zapp:2008gi}. The most modern jet quenching analyses take into account the different regimes of parton virtuality as described in ~\cite{JETSCAPE:2021ehl}. The calculations have as leading input parameter a jet transport coefficient $\hat{\rm q}$ that is determined by the differential mean squared momentum transfer $\langle k_t^2\rangle$ between jet parton and the QGP constituents with respect to the length traversed, i.e. $\hat{\rm q} = d\langle k_t^2\rangle/dL$. 

The recent analysis by the JETSCAPE collaboration~\cite{JETSCAPE:2021ehl} uses data on inclusive hadron suppression from central Au-Au collisions at RHIC and Pb-Pb collisions at LHC, applying a Bayesian parameter estimation to determine the temperature dependence of the dimensionless, renormalized jet transport parameter $\hat{\rm q}/T^3$. The calculations are based on two different models for parton energy loss, called MATTER and LBT, to effectively cover the large range of parton virtualities. A switch-over between the virtuality-ordered splitting dominated regime and the time-ordered transport dominated regime happens at low virtualities of $Q_0 = 2-2.7$ GeV. The results are shown in Fig.~\ref{fig:qhat}. Note that the plot is for a parton momentum of 100 GeV/c, but as demonstrated in ~\cite{JETSCAPE:2021ehl} the momentum dependence is rather mild. To put the results into context, a value of $\hat{\rm q}/T^3 = 4$ implies that, at temperature T = 0.4 GeV, $\hat{\rm q} \approx 1.3 \, \rm{GeV}^2/\rm{fm}$. This value should be compared to what was determined for parton energy loss in cold nuclear matter. Analysis of data for deep inelastic scattering off large nuclei~\cite{Deng:2009ncl} yielded a value of $\hat{\rm q} = 0.024 \pm 0.008 \, \rm{GeV}^2/fm$. A global analysis of the jet transport coefficient for cold nuclear matter was performed recently in ~\cite{Ru:2019qvz}. These authors obtain values of $\hat{\rm q} < 0.03 \, \rm{GeV}^2/fm$ over a wide range of (x$_B,\rm{Q}^2$) values (here, x$_B$ is the Bjorken x parameter). We conclude that, for high energy partons, the stopping power of a QGP formed at RHIC or LHC energy is increased by more than a factor of 40 compared to that for cold nuclear matter. The dramatic jet quenching observed experimentally as displayed in Fig.~\ref{fig:jet_RAA} finds its natural explanation in the large values of the transport coefficient $\hat{\rm q}$ of the QGP.

Direct experimental access to the QCD phase diagram is obtained from the measurement of the yields of hadrons produced in (central) high energy nuclear collisions. Analysis of these data in terms of the Statistical Hadronization Model (SHM), see ~\cite{Andronic:2017pug} and refs. given there, established that, at hadronization, the fireball formed in the collision is very close to a state in full (hadro-)chemical equilibrium. 

\begin{figure}[!htb]
    \centering
    \includegraphics[width=.82\linewidth,clip=true]{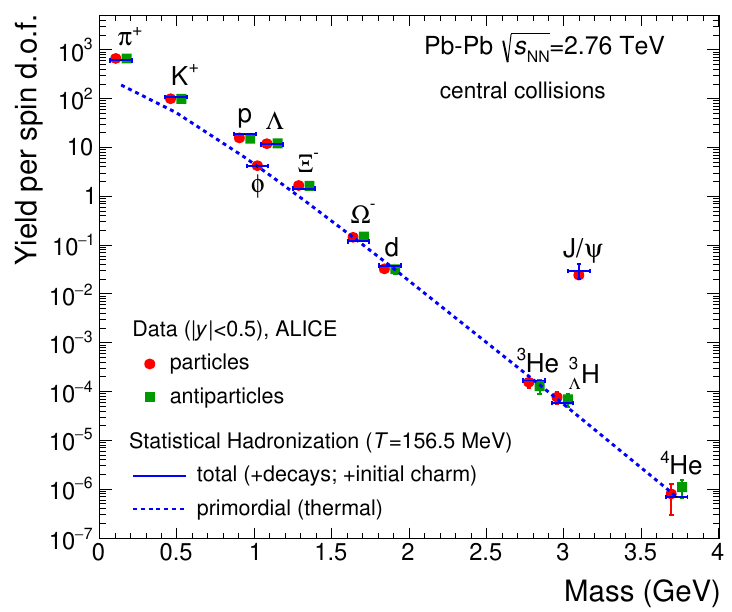}
    \caption{Primordial and total (anti-)particle yields, normalized to the spin degeneracy, as calculated within the  SHMc~\cite{Andronic:2017pug}.}
    \label{fig:HRG}
\end{figure}

The essential idea in the SHM is to approximate the partition function of the system by that of an ideal gas composed of all stable hadrons and resonances, hence also referred to as the Hadron Resonance Gas (HRG) model, see ~\cite{Andronic:2017pug}. From this partition function one can calculate the first moments (mean values) of densities of hadrons as a function of a pair of thermodynamic parameters, the temperature $T_{chem}$ and the baryon chemical potential $\mu_B$ at chemical freeze-out. To go beyond the ideal gas approximation, attractive and repulsive interactions between hadrons can be taken into account in the S-matrix formulation of statistical mechanics~\cite{Dashen:1969ep} by including the first term in the virial expansion. Ideally, the relevant coefficients are obtained from measured phase shifts. For the pion-nucleon interaction this was implemented in~\cite{Lo:2017lym} and the proton yield for LHC energy was corrected accordingly~\cite{Andronic:2018qqt}. The predictions of the SHM for hadron yields are compared to experimental data at LHC energy for $T_{chem}  = 156.5$ MeV in Fig.~\ref{fig:HRG}. The agreement is excellent for the yields of all measured hadrons, nuclei and hyper-nuclei and their anti-particles, with yields varying over 9 orders of magnitude. Remarkably, the description works equally well for loosely bound states. This has led to the conjecture of hadronization into compact multi-quark bags with the right quantum numbers evolving into the final nuclear wave functions in accordance with quantum mechanics~\cite{Andronic:2017pug}.

The values of the hadro-chemical freeze-out parameters at lower collisions energies are similarly obtained by fitting the SHM results to the measured hadron yields. The extracted freeze-out parameters $T_{chem}$ and $\mu_{B}$~\cite{Andronic:2017pug, STAR:2017sal} are presented as red symbols in the QCD phase diagram shown in Fig.~\ref{fig:phase_diagram}. Also included is a  freeze-out point from the HADES collaboration in Au-Au collisions at $\sqrt{s_{\mathrm{NN}}}$ $\approx$ 2.4 GeV~\cite{HADES_FO}. They can be compared to the crossover chiral phase transition line as computed in lQCD (blue band). From LHC energies down to about $\sqrt{s_{\mathrm{NN}}}$ = 12 GeV, i.e., over the entire range covered by lQCD, there is a remarkable agreement between $T_{chem}$ and the pseudo-critical temperature for the chiral cross over transition $T_{pc}$. We note that, along this phase boundary, the energy density computed (for 2 quark flavors) from the values of $T_{chem}$ and $\mu_B$ exhibits a nearly constant value of $\epsilon_{crit} \approx 0.46$ GeV/fm$^3$. 

The finding that the hadro-chemical freeze-out temperature is very close to $T_{pc}$ has a fundamental consequence: because of the very rapid temperature and density change across the phase transition and the resulting low hadron densities in the fireball combined with its size, the produced hadrons cease to interact inelastically within a narrow temperature interval~\cite{Braun-Munzinger:2003htr} after hadron formation. 

This is very different from particle freeze-out in the early universe where for temperatures $T > 10$ MeV even the mean free path for neutrinos is much smaller than its size, see section 22.3 of ~\cite{Workman:2022ynf}.

For large values of baryon chemical potential, experimental data for hadron-chemical freeze-out exist but the phase structure of strongly interacting matter remains uncertain; various model calculations suggest the appearance of a line of first order phase transition, which in combination with the crossover transition at smaller values of $\mu_{B}$, would imply the existence of a critical end point (CEP) in the QCD phase diagram as indicated in Fig.~\ref{fig:phase_diagram}. The experimental discovery of the CEP would mark a major break-through in our understanding of the QCD phase structure. The location of the CEP is most likely in the region $\mu_B > 470$ MeV, based mostly on results from lQCD.  Searching for the CEP is the subject of a very active research program,  at RHIC and the  future FAIR facility at GSI. The importance of this research is underlined by the realization that we have currently no experimental evidence for the order of the chiral phase transition at any value of baryon chemical potential.

\begin{figure}[!htb]
    \centering
    \includegraphics[width=.82\linewidth,clip=true]{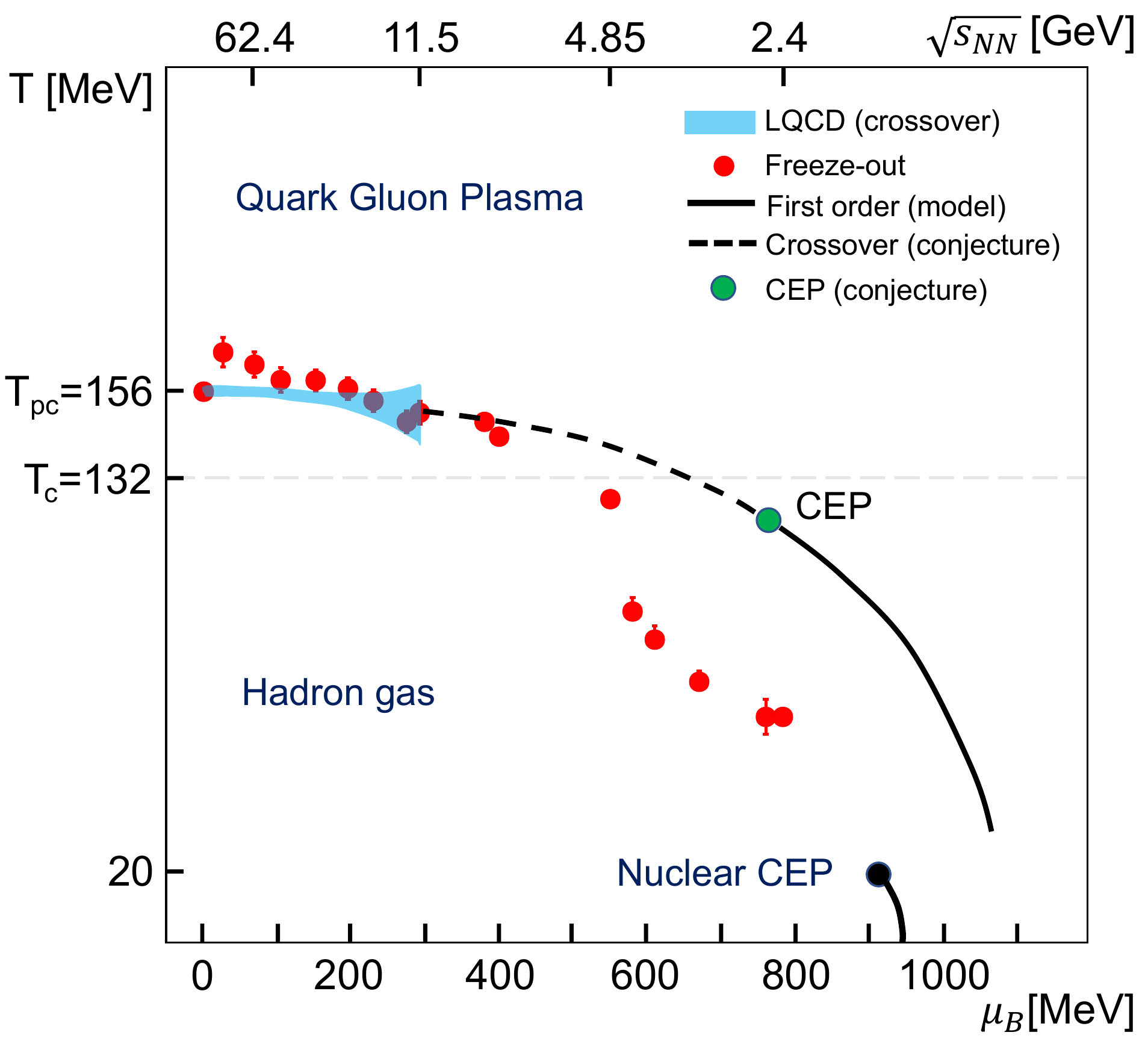}
    \caption{Phase diagram of strongly interacting matter. The red symbols correspond to chemical-freezeout parameters, temperature  $T_{chem}$ and baryon chemical potential $\mu_{B}$ determined from experimental hadron yields~\cite{Andronic:2017pug, STAR:2017sal, HADES_FO}. The blue band represents the results of lQCD computations of the chiral phase boundary~\cite{HotQCD:2018pds,Borsanyi:2020fev}. Also shown are a conjectured line of first order phase transition with a critical end point as well as the nuclear liquid-gas phase boundary.}
    \label{fig:phase_diagram}
\end{figure}

Important further information on the phase structure of QCD matter is expected by measuring, in addition to the first moments of hadron production data, also higher moments as such data can be directly connected to the QCD partition function via conserved charge number susceptibilities in the Grand Canonical Ensemble (GCE)~\cite{Bzdak:2016sxg,Bzdak:2019pkr}. For a thermal system of volume $V$ and temperature $T$ the susceptibilities in the GCE are defined as the coefficients in the Maclaurin series of the reduced pressure $\hat{P}=P(T, V, \vec{\mu})/T^{4}$
\begin{equation}
\label{eq-chi}
    \chi_{n}^q \equiv \frac{\partial^{n}\hat{P}}{\partial \hat{\mu}_{q}^n} = \frac{1}{VT^{3}}\frac{\partial^{n} ln Z(V, T, \vec{\mu})}{\partial \hat{\mu}_{q}^{n} } = \frac{\kappa_n(N_q)}{VT^3},
\end{equation}
where $\vec{\mu} = \{\mu_{B}, \mu_{Q}, \mu_{S}\}$ is the chemical potential vector that is introduced to conserve, on average, baryon number, electric charge and strangeness. Here, $\hat{\mu}_{q}=\mu_{q}/T$ is the reduced chemical potential for the conserved charges $q \in \{B, Q, S\}$. The partition function  $Z(V, T, \vec{\mu})$ encodes the Equation of State (EoS) of the system under consideration. Eq.~\ref{eq-chi} establishes a direct link between susceptibilities and fluctuations of conserved charge numbers. By measuring cumulants $\kappa_{n}(N_q)$ of net-charge number ($N_{q}$) distributions  one can, using Eq.~\ref{eq-chi}, further probe and quantify the nature of the QCD phase transition. 

Important at this point is to define a non-critical baseline, which is done by using the ideal gas EoS, extended such as to account for event-by-event charge conservation and correlations in rapidity space~\cite{Braun-Munzinger:2018yru, Braun-Munzinger:2020jbk, Braun-Munzinger:2019yxj}, see also ~\cite{Vovchenko:2020gne}. In addition, non-critical contributions arising, e.g., from fluctuations of wounded nucleons~\cite{Braun-Munzinger:2016yjz, Skokov:2012ds} need to be corrected for. Deviations from this non-critical baseline, for example leading to negative values of $\kappa_{6}$ for net-baryons would arise due to the closeness of the cross over transition to the O(4) 2nd order critical phase transition for vanishing light quark masses ~\cite{Bazavov:2020bjn}.

\begin{figure}[!htb]
    \centering
    \includegraphics[width=.8\linewidth,clip=true]{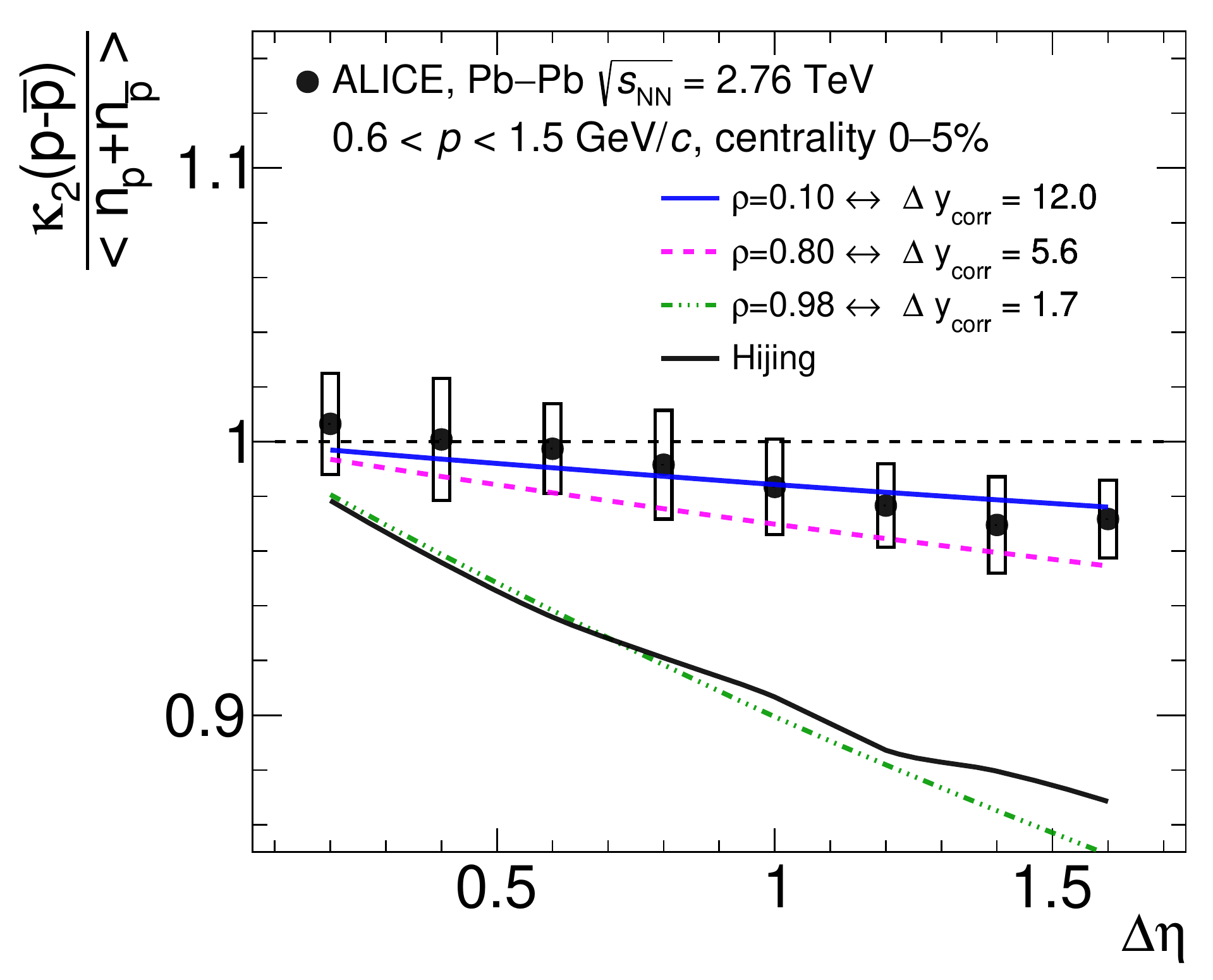}
    \caption{Scaled second order cumulants of the net-proton distribution as a function of the pseudo-rapidity acceptance measured by the ALICE experiment (black symbols)~\cite{ALICE:2019nbs}. The colored lines correspond to calculations accounting for baryon number conservation with different correlation length in rapidity space~\cite{Braun-Munzinger:2019yxj}. The results of the HIJING event generator are presented with the black solid line.}
    \label{fig:AliceKappa2}
\end{figure}

In Fig.~\ref{fig:AliceKappa2} the ALICE results  on the normalized second order cumulants of net-proton distributions are presented as function of the experimental acceptance. The acceptance is quantified via the pseudo-rapidity coverage around mid-rapidity $\Delta \eta$~\cite{Rustamov:2017lio, ALICE:2019nbs, Rustamov:2020ekv}. The measured cumulant values approach unity at small values of $\Delta \eta$, essentially driven by small number Poisson statistics. With increasing acceptance, the data progressively decrease from unity. For small but finite acceptance the decrease can be fully ac\-coun\-ted for by overall baryon number conservation in full phase space. Hence, after correcting for baryon number conservation, the experimental data would be consistent with unity over the range of the experimental acceptance. 

This observation has three important consequences. (i) It shows that, up to second order,  cumulants of the baryon number distribution functions follow a poissonian distribution, a posteriori  justifying the assumptions underlying the  construction of the partition function used in the SHM. (ii) This is the first experimental verification of lQCD results which also predict unity for the second order scaled cumulants of baryon distributions. (iii) Compared to the different calculations, the data imply long range correlations in rapidity space, calling into question the baryon production mechanism implemented in string fragmentation models. Indeed, the results from the HIJING event generator based on the Lund String Fragmentation model shown in Fig.~\ref{fig:AliceKappa2}, due to the typical correlation over about one unit of rapidity, grossly overpredict the suppression due to baryon number conservation~\cite{Rustamov:2022hdi}. 

\begin{figure}[!htb]
    \centering
    \includegraphics[width=.8\linewidth,clip=true]{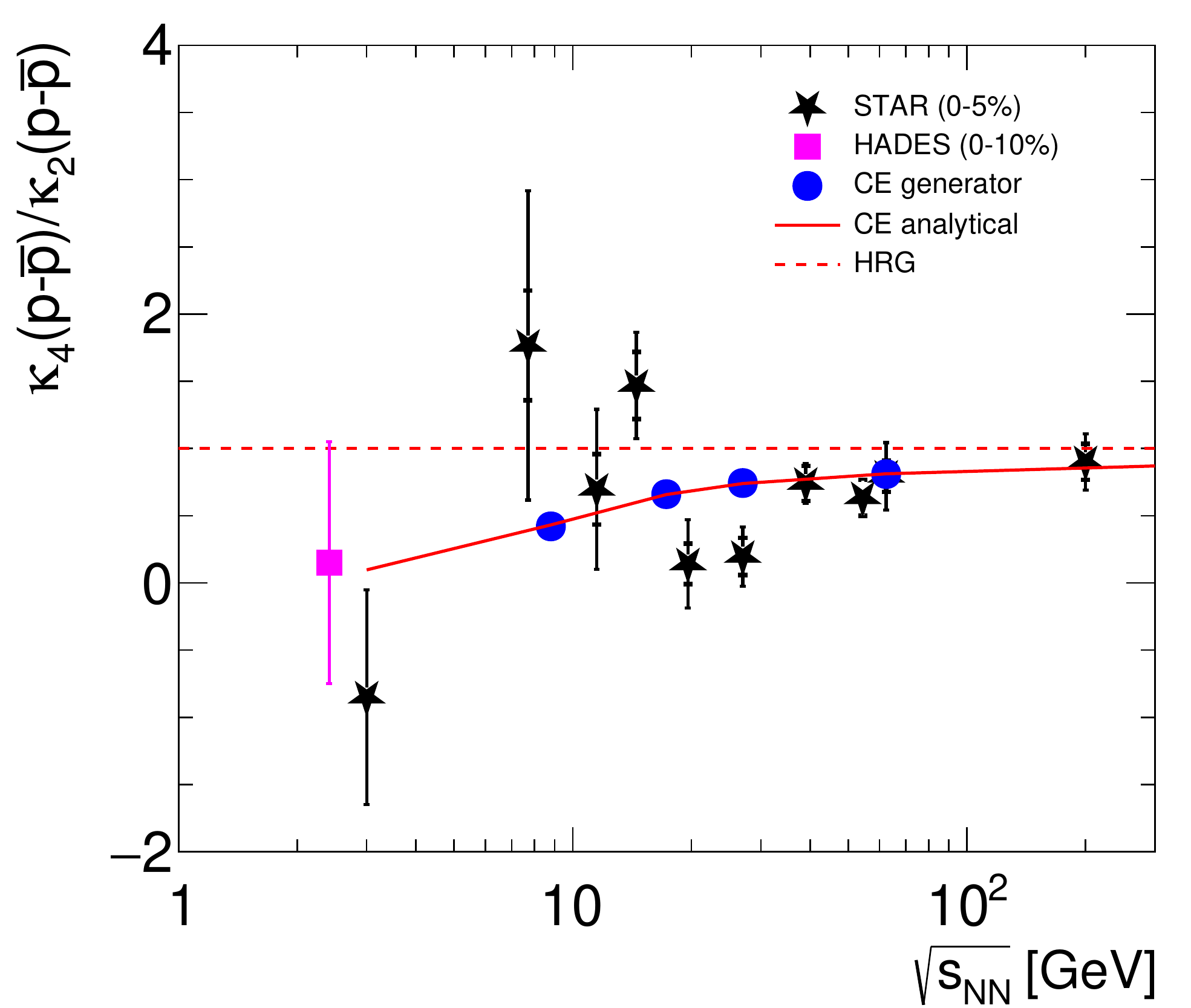}
    \caption{Collision energy dependence of the fourth to second order cumulants of net-proton distributions as measured by experiments. The STAR data are for $|y| < 0.5$ and $p_t$ = 0.4 - 2 GeV/c, the HADES data for $|y| < 0.4$ and $p_t$ = 0.4 - 1.6 GeV/c. The non-critical baseline induced by global baryon number conservation is indicated by the blue circles and the red line.}
    \label{fig:CP}
\end{figure}

Contrary to the detailed predictions for signals in the cross-over region of the transition covered by lQCD, no quantitative signals are available for the existence of a possible critical end point in the phase diagram. All predicted signals are of  generic nature and mostly based on searching for non-monotonic behavior in the excitation function of fourth order cumulants of, e.g., net-protons~\cite{Stephanov:2011pb}. A compilation of the respective measurements~\cite{HADES:2020wpc, STAR:2020tga} is presented in Fig.~\ref{fig:CP}. The search for non-monotonic behaviour needs a starting point. In Fig.~\ref{fig:CP} two possibilities are presented, one corresponding to calculations in HRG within GCE (dashed line at unity) and the other the non-critical baseline introduced above where baryon number conservation is explicitly accounted for (red solid line or blue symbols). With respect to unity the data indeed exhibit an indication for non-monotonic behaviour with a significance corresponding to 3.1 standard deviations~\cite{STAR:2020tga}. However, a significant part of this deviation from unity is induced by non-critical effects, such as baryon number conservation. Therefore, one must search for non-monotonic behaviour with respect to the red solid line. Analysis of the data shows that there are no statistically significant deviations from a statistical ensemble with baryon number conservation, i.e, within the current precision of the data there is not yet evidence for the presence of a critical end point~\cite{Braun-Munzinger:2018yru, Braun-Munzinger:2020jbk}. The analysis of fourth order cumulants from a much higher statistics data set has just started and will be essential for a possible discovery of the critical point.

The current status on experimental verification of the nature of the chiral cross-over transition at vanishing or moderate $\mu_{B}$ is still rather open. Within QCD inspired model calculations ~\cite{Friman:2011pf, Almasi:2017bhq}, based on O(4) scaling functions the predicted sixth order cumulants for net-baryon distributions exhibit negative values at $T_{pc}$ due to a singular term in the pressure. Similarly, the sixth order susceptibilities of baryon number resulting from lQCD calculations are also negative~\cite{Bazavov:2020bjn, Borsanyi:2018grb} and this sign change (relative to the HRG baseline in GCE) has been linked to the critical component in the pressure present as a residue of the 2nd order chiral phase transition for vanishing (u,d) quark masses, due to the smallness of the physical masses. First experimental results on sixth order net-proton cumulants were reported by the STAR collaboration~\cite{STAR:2021rls} for Au--Au collisions, albeit with sizeable statistical uncertainties since the data analysis to determine high order cumulants is extremely statistics hungry. Qualitatively, the STAR results at $\sqrt{s_{\mathrm{NN}}}$ = 200 GeV are indeed consistent with the expectations for the crossover transition. At the same time, the experimentally measured energy dependence of $\kappa_{6}$~\cite{STAR:2021rls} is at odds with both model and lQCD calculations. For a quantitative conclusion, in any case, the effects of baryon number conservation~\cite{Braun-Munzinger:2020jbk} and transformation from net-protons (experiment) to net-baryons (theory)~\cite{Kitazawa:2012at} are still to be performed. So far, experimental insight into the nature of the chiral cross-over transition and the development towards low net-baryon densities remains inconclusive. It can be expected that ongoing and future high statistics measurement campaigns by the STAR and ALICE collaborations will elucidate the situation.

There is now significant experimental information, from relativistic nuclear collisions, not only on the production of hadrons composed of light (u,d,s) quarks, but also of open and hidden charm and beauty hadrons. In particular, there is good evidence, mainly from results obtained at the CERN Large Hadron Collider (LHC) ~\cite{ALICE:2021rxa,Andronic:2021erx,Andronic:2019wva}, that charm quarks reach a large degree of thermal equilibrium, although charm quarks in the system are chemically far out of equilibrium. This is supported by heavy quark diffusion coefficients from lQCD~\cite{Altenkort:2020fgs}. A strong indication for equilibration is the fact that J/$\psi$ mesons participate in the collective, anisotropic hydrodynamic expansion ~\cite{ALICE:2013xna,He:2021zej}.

To microscopically understand the production me\-cha\-nism of charmed hadrons for systems ranging from pp to Pb--Pb, various forms of quark coalescence mo\-dels have been developed ~\cite{Cho:2019lxb,Zhao:2018jlw,ExHIC:2017smd,Zhou:2014kka,Greco:2003vf}. This provides a natural way to study the dependence of production yields on hadron size and, hence, may help to settle the still open question whether the many exotic hadrons that have been observed recently are compact multi-quark states or hadronic molecules (see ~\cite{Aarts:2016hap,Maiani:2022psl} and refs. cited there). Conceptual difficulties with this approach are that energy is not conserved in the coalescence process and that color neutralization at hadronization requires additional assumptions about quark correlations in the QGP~\cite{Song:2021mvc}.

Another approach, named SHMc, has been made possible by the extension of the SHM to also incorporate charm quarks. This was first proposed in ~\cite{Braun-Munzinger:2000csl} and developed further in ~\cite{Andronic:2002pj,Grandchamp:2003uw,Becattini:2005hb,Andronic:2006ky,Andronic:2017pug,Andronic:2021erx} to include all hadrons with hidden and open charm. The key idea is based on the recognition that, contrary to what happens in the (u,d,s) sector, the heavy (mass $\sim$ 1.2 GeV) charm quarks are not thermally produced. Rather, production takes place in initial hard collisions. The produced charm quarks then thermalize in the hot fireball, but the total number of charm quarks is conserved during the evolution of the fireball~\cite{Andronic:2006ky} since charm quark annihilation is very small. In essence, this implies that charm quarks can be treated like impurities. Their thermal description then requires the introduction of a charm fugacity $g_c$~\cite{Braun-Munzinger:2000csl,Andronic:2021erx}. The value of $g_c$ is not a free parameter but experimentally determined by measurement of the total charm cross section. For central Pb--Pb collisions at LHC energy, $g_c \approx 30$~\cite{Andronic:2021erx}. The charmed hadrons are, in the SHMc, all formed at the phase boundary, i.e. at hadronization, in the same way as all (u,d,s) hadrons. 

In Fig.~\ref{fig:HRG} it can be seen that, with that choice, the measured yield for J/$\psi$ mesons is very well reproduced, the uncertainty in the prediction is mainly caused by the uncertainty in the total charm cross section in Pb--Pb collisions. We note here that, because of the formation from deconfined charm quarks at the phase boundary, charmonia are unbound inside the QGP but their final yield exhibits enhancement compared to expectations using collision scaling from pp collisions, contrary to the original predictions based on ~\cite{Matsui:1986dk}. For a detailed discussion see ~\cite{Andronic:2017pug}.

\begin{figure}[!htb]
    \centering
    \includegraphics[width=.8\linewidth,clip=true]{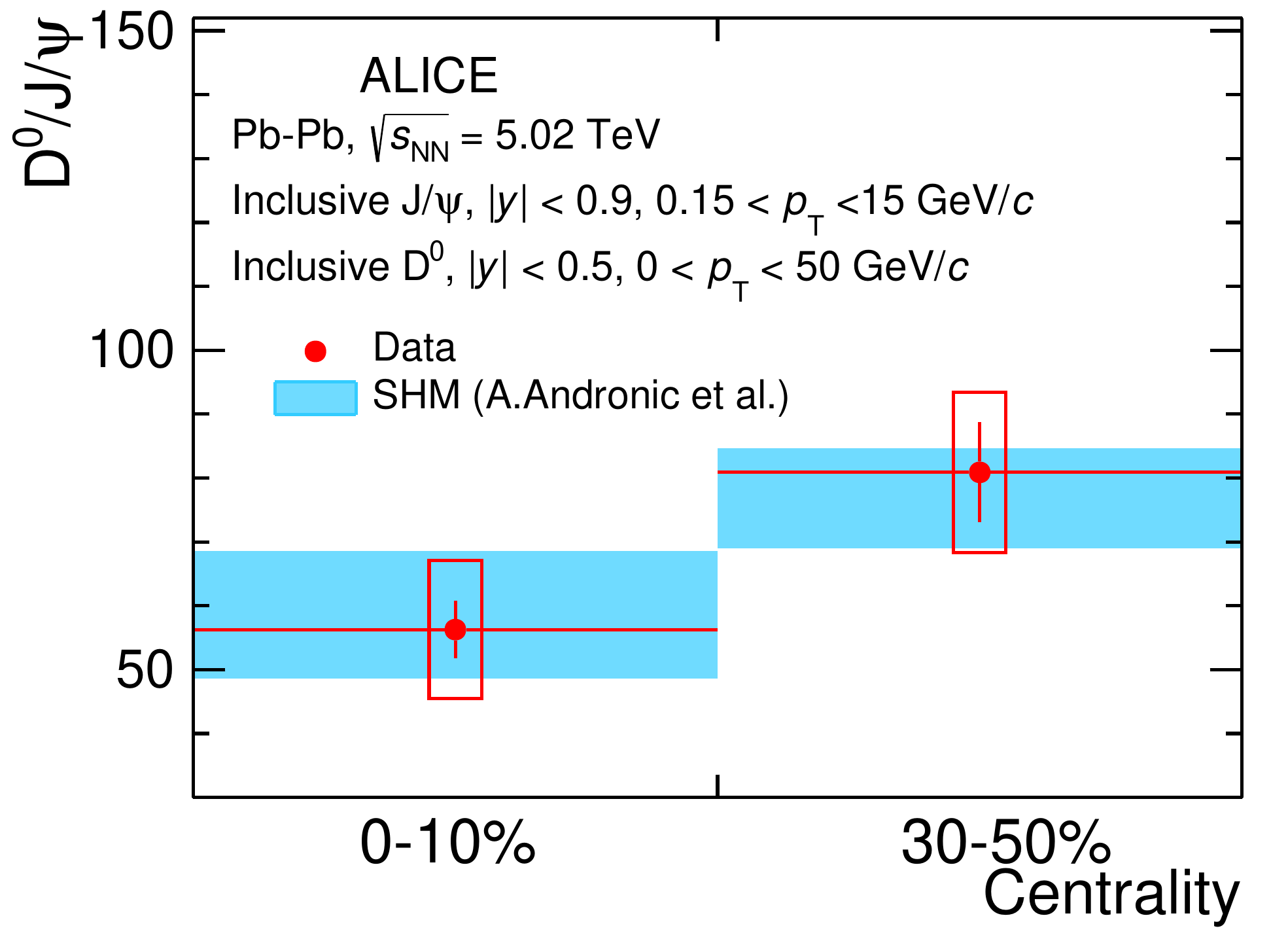}
    \caption{$D^{0}$ to J/$\psi$ yield ratio measured in Pb--Pb collisions at the LHC and predicted by the Statistical Hadronization Model with charm SHMc. Figure from~\cite{ALICE:2022new}.}
    \label{fig:DotoJpsi}
\end{figure}

For the description of yields of charmonia, feeding from excited charmonia is very small because of their strong Boltzmann suppression. For open charm mesons and baryons, this is not the case and feeding from excited $D^*$ and $\Lambda_c^*$ is an essential ingredient for the description of open charm hadrons ~\cite{Andronic:2021erx}. Even though the experimental delineation of the mass spectrum of excited open charm mesons and baryons is currently far from complete, the prediction of yields for D-mesons and $\Lambda_c$ baryons compares very well with the measurements\footnote{For $\Lambda_c$ baryons on has to augment the currently measured charm baryon spectrum with additional states to achieve complete agreement with experimental data~\cite{Andronic:2021erx}.}, both concerning transverse momentum and centrality dependence.

A particularly transparent way to look at the data for Pb--Pb collisions is obtained by analyzing the centrality dependence of the yield ratio $D^0/(J/\psi)$ and comparing the results to the predictions of the SHMc. Recently, both the $D^0$ and $J/\psi$ production cross sections have been well measured down to $p_t$ = 0. The yield ratio $D^0/(J/\psi)$ is reproduced with very good precision for both measured centralities, as demonstrated in Fig.~\ref{fig:DotoJpsi}. This result lends strong support to the assumption that open and hidden charm states are both produced by statistical hadronization at the phase boundary. A more extensive comparison between SHMc and data for open charm hadrons is shown in~\cite{Andronic:2021erx}.

From the successful comparison of measured yields for the production of (u,d,s) as well as open and hidden charm hadrons obtained from the SHM or SHMc with essentially only the temperature as a free parameter at LHC energies, one may draw a number of important conclusions. 
\begin{itemize}
\item 
First, we note that hadron production in relativistic nuclear collisions  is described quantitatively by the chemical freeze-out parameters  ($T_{chem}, \mu_{B}$). Note that the fireball volume appearing in the partition function is determined by normalization to the measured number of charged particles. At least for energies $\sqrt{s_{NN}}\geq$ 10 GeV these freeze-out parameters agree with good precision with the results from lQCD for the location of the chiral cross over transition. Under these conditions, hadronization is independent of particle species and only dependent on the values of $T$ and $\mu_B$ at the phase boundary. At LHC energy, the chemical potential vanishes, and only $T = T_{pc}$ is needed to describe hadronization.
      
\item
The mechanism implemented in the SHMc for the production of charmed hadrons implies that these particles are produced from uncorrelated, thermalized charm quarks as is expected for a strongly coupled, deconfined QGP (see also the discussion in ~\cite{Andronic:2021erx}).  At LHC energy, where chemical freeze-out takes place  for central Pb--Pb collisions in a volume per unit rapidity of $V \approx 4000 $ fm$^3$, this implies that charm quarks can travel over linear distances of order 10 fm (see ~\cite{Andronic:2017pug,Andronic:2021erx} for more detail). 
\end{itemize}

One may ask whether there is a possible contribution to the production of charmed hadrons (in particular of J/$\psi$) from the hadronic phase. At the phase boundary, assembly of J/$\psi$ from deconfined charm quarks or from all possible charmed hadrons is indistinguishable, as discussed in detail in~\cite{Andronic:2017pug}. In fact, in~\cite{Braun-Munzinger:2003htr} it was demonstrated that  multi-hadron collisions lead to very rapid thermal population, while within very few MeV below the phase boundary, the system falls out of equilibrium. Both is driven by the rapid drop of entropy and thereby particle density in the vicinity of $T_{pc}$. In the confined hadronic phase, i.e. for temperatures lower than $T_{pc}$, the hadron gas is off-equilibrium, and any calculation via reactions of the type $D\bar D^* \leftrightarrow n\pi \rm{J}/\psi$ has to implement the back-reaction~\cite{Rapp:2000gy}. Since predictions with the SHMc agree very well with the data for J/$\psi$ production at an accuracy of about 10\%, and since any possible hadronic contribution has to be added to the SHMc value, we estimate any contribution to J/$\psi$ production from the confined phase to be less than 10\%.

Future measurement campaigns at the LHC will yield detailed information on the production cross sections of hadrons with multiple charm quarks as well as excited charmonia. The predictions from the SHMc for the relevant cross sections exhibit a rather dramatic hie\-rarchy of enhancements~\cite{Andronic:2021erx} for such processes. Expe\-ri\-men\-tal tests of these predictions would lead to a fundamental understanding of confinement/deconfinement and hadronization.

\section{Acknowledgements} 
We acknowledge continued and long-term collaboration with Anton Andronic and Krzysztof Redlich on many of the topics described in this contribution. This work is supported by the DFG Collaborative Research Centre ”SFB 1225 (ISOQUANT)”.

\bibliography{PBM_AR_JS.bib}{}

%\printbibliography
\end{document}